\documentclass[aps,prl,twocolumn,superscriptaddress,floatfix,letter]{revtex4}

\usepackage{amsmath,amssymb,amsthm}
\usepackage{times}
\usepackage{graphicx}

\newcommand{\WFini}{\vert \psi(0)\rangle}
\newcommand{\WFtime}{\vert \psi(t)\rangle}
\newcommand{\Eigenv}{\vert\Psi_\alpha\rangle}
\newcommand{\CEigenv}{\langle\Psi_{\alpha'}\vert}
\newcommand{\AEigenv}{\langle\Psi_\alpha\vert}
\newcommand{\Proj}{\vert\Psi_\alpha\rangle \langle \Psi_\alpha \vert}
\newcommand{\emphas}[1]{\textit{#1}}

\newcommand{\leftindex}{\alpha}
\newcommand{\rightindex}{\beta}
\newcommand{\Eigenvleft}{\vert\Psi_\leftindex\rangle}

\newcommand{\WhetherDisplayFigure}[1]{\ifthenelse{\boolean{DontDisplayGraphics}}{}{#1}}
\newcommand{\ignore}[1]{}

\begin{document}

\title{Thermalization and its mechanism for generic isolated
quantum systems\footnote{Published in 
Nature {\bf 452}, 854-858 (17 April 2008); 10.1038/nature06838. 
http://www.nature.com/nature/journal/v452/n7189/abs/nature06838.html}}

\author{Marcos Rigol}
\affiliation{Department of Physics \& Astronomy, University of Southern California, Los Angeles, CA 90089, USA}
\affiliation{Department of Physics, University of Massachusetts Boston, Boston, MA 02125, USA}
\author{Vanja Dunjko}
\affiliation{Department of Physics \& Astronomy, University of Southern California, Los Angeles, CA 90089, USA}
\affiliation{Department of Physics, University of Massachusetts Boston, Boston, MA 02125, USA}
\author{Maxim Olshanii}
\affiliation{Department of Physics, University of Massachusetts Boston, Boston, MA 02125, USA}

\begin{abstract}
Time dynamics of isolated many-body quantum systems has long been an elusive subject. Very recently, however, meaningful experimental studies of the problem have finally become possible \cite{kinoshita06,hofferberth07}, stimulating theoretical interest as well \cite{rigol07,kollath07,manmana07,burkov07,calabrese07}. Progress in this field is perhaps most urgently needed in the foundations of quantum statistical mechanics. This is so because in generic isolated systems, one expects \cite{sengupta04,berges04} nonequilibrium dynamics on its own to result in thermalization: a relaxation to states where the values of macroscopic quantities are stationary, universal with respect to widely differing initial conditions, and predictable through the time-tested recipe of statistical mechanics. However, it is not obvious what feature of many-body quantum mechanics makes quantum thermalization possible, in a sense analogous to that in which dynamical chaos makes classical thermalization possible \cite{gallavotti99}. For example, dynamical chaos itself cannot occur in an isolated quantum system, where time evolution is linear and the spectrum is discrete \cite{krylov79}. Underscoring that new rules could apply in this case, some recent studies even suggested that statistical mechanics may give wrong predictions for the outcomes of relaxation in such systems \cite{kollath07,manmana07}. Here we demonstrate that an isolated generic quantum many-body system does in fact relax to a state well-described by the standard statistical mechanical prescription. Moreover, we show that time evolution itself plays a merely auxiliary role in relaxation and that thermalization happens instead at the level of individual eigenstates, as first proposed by J. M. Deutsch \cite{deutsch91} and M. Srednicki \cite{srednicki94}. A striking consequence of this \emphas{eigenstate thermalization} scenario, confirmed below for our system, is that the knowledge of a \emphas{single} many-body eigenstate suffices to compute thermal averages---any eigenstate in the microcanonical energy window will do, as they all give the same result.
\end{abstract}

\maketitle

If we pierce an inflated balloon inside a vacuum chamber, very soon we find the released air uniformly filling the enclosure and the air molecules attaining the Maxwell velocity distribution whose width depends only on their total number and energy. Different balloon shapes, placements, or piercing points all lead to the same spatial and velocity distributions. Classical physics explains this \textit{thermodynamical universality} as follows \cite{gallavotti99}: almost all particle trajectories quickly begin looking alike, even if their initial points are very different, because nonlinear equations drive them to explore ergodically the constant-energy manifold, covering it uniformly with respect to precisely the microcanonical measure. However, if the system possesses further conserved quantities \textit{functionally independent} from the Hamiltonian and each other, then time evolution is confined to a highly restricted hypersurface of the energy manifold. Hence, microcanonical predictions fail and the system does not thermalize.

On the other hand, in isolated quantum systems not only is dynamical chaos absent due to the linearity of time evolution and the discreteness of spectra \cite{krylov79}, but it is also not clear under what conditions conserved quantities provide independent constraints on relaxation dynamics. To begin with, any operator commuting with a generic and thus nondegenerate Hamiltonian is functionally dependent on it \cite{sutherland04}, implying that the conservation of energy is the only independent constrain. On the other hand, even when operators are functionally dependent, their expectation values---considered as functionals of states---generally are not: for example, two states may have the same mean energies but different mean square-energies. For nondegenerate Hamiltonians a maximal set of constants of motion with functionally independent expectation values is as large as the dimension of the Hilbert space; examples include the projectors $\widehat{P}_\alpha=\Proj$ to the energy eigenstates \cite{sutherland04} and the integer powers of the Hamiltonian \cite{manmana07}.

The current numerical and analytic evidence from integrable systems suggests that there exists a minimal set of independent constraints whose size is much smaller than the dimension of the Hilbert space but may still be much greater than one. In our previous work \cite{rigol07} (with V. Yurovsky) we showed that an integrable isolated one-dimensional system of lattice hard-core bosons relaxes to an equilibrium characterized not by the usual but by a \textit{generalized} Gibbs ensemble. Instead of just the energy, the Gibbs exponent contained a linear combination of conserved quantities---the occupations of the eigenstates of the corresponding Jordan-Wigner fermions---whose number was still only a tiny fraction of the dimension of the Hilbert space. Yet this ensemble works, while the usual one does not, for a wide variety of initial conditions \cite{rigol06} as well as for a fermionic system \cite{cazalilla06}; it also explains a recent experimental result, the absence of thermalization in the Tonks-Girardeau gas \cite{kinoshita06}. Thus, while at least some constraints beyond the conservation of energy must be kept, it turns out one needs only a relatively limited number of additional conserved quantities with functionally independent expectation values; adding still further ones is redundant.

Since it is not clear which sets of conserved quantities---and some are always present---constrain relaxation and which do not, it becomes even more urgent to determine whether isolated generic quantum systems relax to the usual thermal state. The theoretical attention to this question has in fact been increasing recently, because of the high levels of isolation \cite{kinoshita06,hofferberth07,jin96} and control \cite{greiner02,mandel03} possible in experiments with ultracold quantum gases. However, despite numerous studies of specific models there is not yet consensus on how or even if relaxation to the usual thermal values occurs for nonintegrable systems \cite{calabrese07}. Common wisdom says that it does \cite{sengupta04,berges04}, but some recent numerical results suggest otherwise, either under certain conditions \cite{kollath07} or in general \cite{manmana07}.


\begin{figure}[!t]
\begin{center}
\includegraphics[scale=.27]{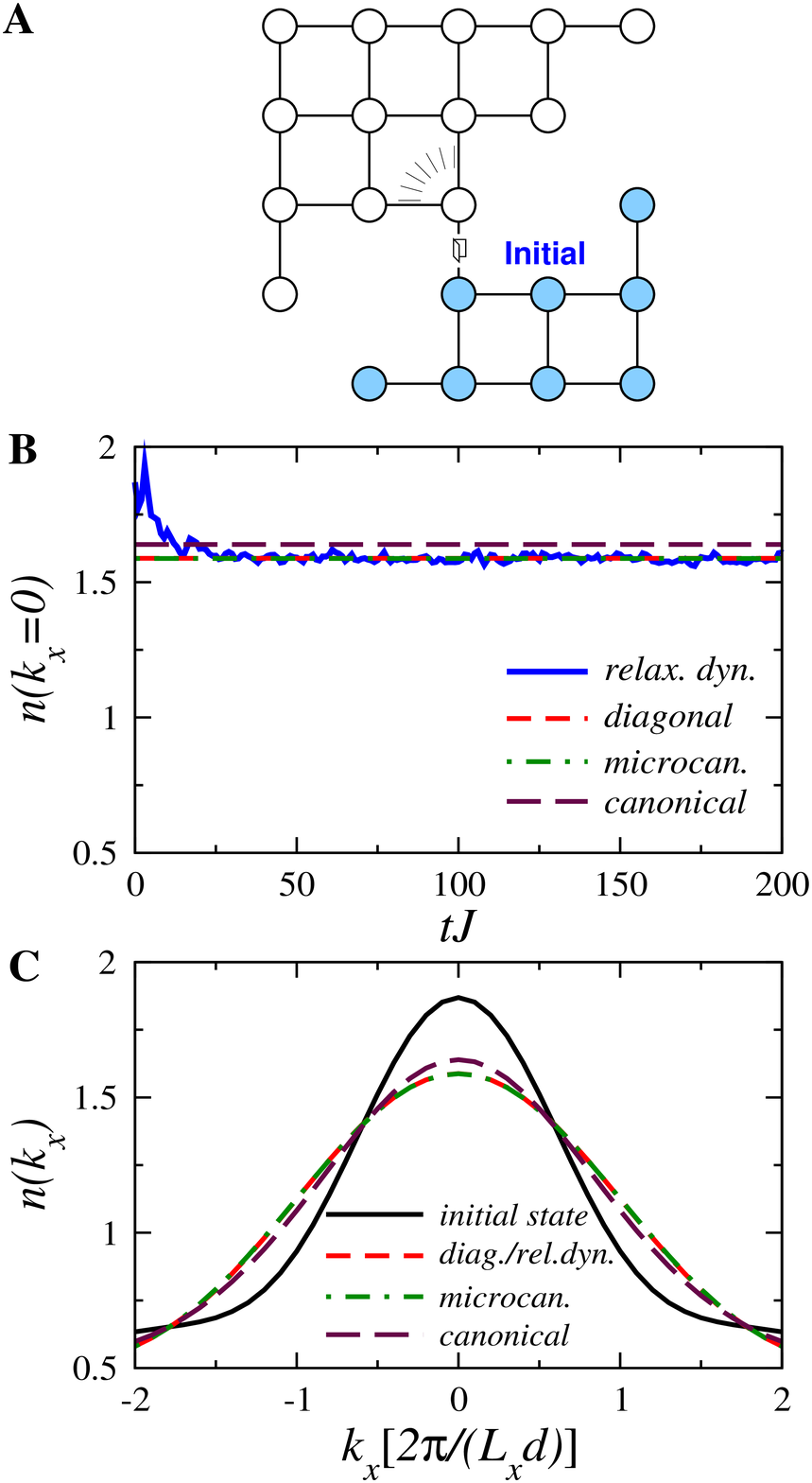}
\end{center}
\vspace{-0.7cm}
\caption{  \label{fig1}
\textbf{Relaxation dynamics}.
\textbf{a}, Two-dimensional lattice on which five hard-core bosons propagate in time. The bosons are initially prepared in the ground state of the sub-lattice in the lower-right corner and released through the indicated link.
\textbf{b}, The corresponding relaxation dynamics of the marginal momentum distribution center [$n(k_{x}=0)$] compared with the predictions of the three ensembles. In the microcanonical case, we averaged over all eigenstates whose energies lie within a narrow window (see Supplementary Discussion) $[E_0-\Delta E, E_0+\Delta E]$, where $E_0\equiv\langle \psi(0) | \widehat{H} | \psi(0) \rangle=-5.06 J$, $\Delta E=0.1J$, and $J$ is the hopping parameter. The canonical ensemble temperature is $k_{\textrm{\scriptsize B}}T=1.87 J$, where $k_{\textrm{\scriptsize B}}$ is the Boltzmann constant, so that the ensemble prediction for the energy is $E_0$.
\textbf{c}, Full momentum distribution function in the initial state, after relaxation, and in the different ensembles. Here $d$ is the lattice constant and $L_{x}=5$ the lattice width.
}
\end{figure}

In order to study relaxation of an isolated quantum systems, we considered the time evolution of five hard-core bosons with additional weak nearest-neighbor repulsions, on a 21-site two-dimensional lattice, initially confined to a portion of the lattice and prepared in their ground state there. Figure \ref{fig1}a shows the exact geometry (see also Supplementary Discussion); the relaxation dynamics begins when the confinement is lifted. Expanding the initial state wavefunction in the eigenstate basis of the final Hamiltonian $\widehat{H}$ as $\WFini=\sum_\alpha C_{\alpha} \Eigenvleft$, the many-body wavefunction evolves as  $\WFtime =e^{-i\widehat{H} t}\WFini=\sum_\alpha C_{\alpha^{}}^{}  e^{-i E_\alpha t}\Eigenvleft$, where the $E_\alpha$'s are the eigenstate energies. Thus obtaining numerically-exact results for all times required the full diagonalization of the 20,349-dimensional Hamiltonian. The quantum-mechanical mean of any observable $\widehat{A}$ evolves as
\begin{equation}
\langle \widehat{A}(t) \rangle
\equiv\langle \psi(t) | \widehat{A} | \psi(t)  \rangle
=\sum_{\leftindex,\,\rightindex} C_{\leftindex}^{\star} C_{\rightindex}^{}
e^{i(E_{\leftindex}-E_{\rightindex})t}
A_{\leftindex\rightindex}\, ,
\label{timeevolution}
\end{equation}
with $A_{\leftindex\rightindex} =\CEigenv \widehat{A} \Eigenv$. The long-time average of $\langle \widehat{A}(t) \rangle$ is then
\begin{eqnarray}
\overline{\langle \widehat{A} \rangle}
= \sum_{\alpha} |C_{\alpha}|^{2}
A_{\alpha\alpha} \, .
\label{diagonal}
\end{eqnarray}
Note that if the system relaxes at all, it must be to this value. We find it convenient to think of Eq. (\ref{diagonal}) as stating the prediction of a ``diagonal ensemble,'' $|C_{\alpha}|^{2}$ corresponding to the weight $\Eigenvleft$ has in the ensemble. In fact, this ensemble is precisely the generalized Gibbs ensemble introduced in Ref.\ \cite{rigol07} if as integrals of motion one takes all the projection operators $\widehat{P}_\alpha=\Proj$. Using these as constraints on relaxation dynamics, the theory gives $\hat{\rho}_c=\exp{\left( -\sum_{\alpha=1}^D \lambda_\alpha \hat{P}_\alpha\right)}$, with $\lambda_\alpha=-\ln (\left|C_\alpha\right|^{2})$, and $D$ the dimension of the Hilbert space. (Notice, however, that for the integrable system treated in Ref.\ \cite{rigol07}, the generalized Gibbs ensemble was defined using a different, \emphas{minimal} set of independent integrals of motion, whose number was equal to the number of lattice sites $N\ll D$.)

Now if the quantum-mechanical mean of an observable behaves thermally it should settle to the prediction of an appropriate statistical-mechanical ensemble. For our numerical experiments we chose to monitor the marginal momentum distribution along the horizontal axis $n(k_{x})$ and its central component $n(k_{x}=$~$0)$ (see Supplementary Discussion). Figures \ref{fig1}b and \ref{fig1}c demonstrate that both relax to their microcanonical predictions. The diagonal ensemble predictions are indistinct from these, but the canonical ones, although quite close, are not. This is an indication of the relevance of finite size effects, which may be the origin of some of the apparent deviations from thermodynamics seen in the recent numerical studies of Refs.\ \cite{kollath07} and \cite{manmana07}.

The statement that the diagonal and microcanonical ensembles give the same predictions for the relaxed value of $\widehat{A}$ reads
\begin{eqnarray}
\sum_{\alpha} |C_{\alpha}|^{2}
A_{\alpha\alpha} &=&
\langle A \rangle_{\text{microcan.}}(E_{0})\, \nonumber \\
& \stackrel{\text{def.}}{=}&\, 
\frac{1}{\mathcal{N}_{E_{0},\, \Delta E}} \sum_{\substack{\alpha \\ \left|E_{0}-E_{\alpha}\right|<\Delta E}}\,A_{\alpha\alpha}
\, ,
\label{paradox}
\end{eqnarray}
where $E_{0}$ is the mean energy of the initial state, $\Delta E$ is the half-width of an appropriately chosen (see Supplementary Discussion) energy window centered at $E_{0}$, and the normalization $\mathcal{N}_{E_{0},\, \Delta E}$ is the number of energy eigenstates with energies in the window $\left[E_{0}-\Delta E,\, E_{0}+\Delta E\right]$. Thermodynamical universality is evident in this equality: while the left hand side depends on the details of the initial conditions through the set of coefficients $C_{\alpha}$, the right hand side depends only on the total energy, which is the same for many different initial conditions.
Three mechanisms suggest themselves as possible explanations of this universality (assuming the initial state is sufficiently narrow in energy, as is normally the case---see Supplementary Discussion):

(i) Even for eigenstates close in energy, there are large eigenstate-to-eigenstate fluctuations of both the eigenstate expectation values $A_{\alpha\alpha}$ and of the eigenstate occupation numbers $|C_{\alpha}|^{2}$. However, for physically interesting initial conditions, the fluctuations in the two quantities are uncorrelated. A given initial state then performs an unbiased sampling of the distribution of the eigenstate expectation values $A_{\alpha\alpha}$, resulting in Eq. (\ref{paradox}).

(ii) For physically interesting initial conditions, the eigenstate occupation numbers $|C_{\alpha}|^{2}$ practically do not fluctuate at all between eigenstates that are close in energy. Again, Eq. (\ref{paradox}) immediately follows.

(iii) The eigenstate expectation values $A_{\alpha\alpha}$ practically do not fluctuate at all between eigenstates that are close in energy. In that case Eq. (\ref{paradox}) holds for literally \textit{all} initial states narrow in energy.

\begin{figure}[!t]
\begin{center}
\includegraphics[scale=.27]{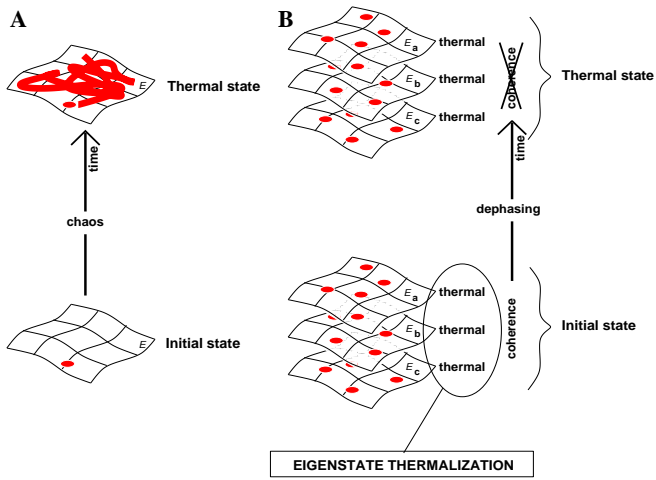}
\end{center}
\vspace{-0.6cm}
\caption {  \label{fig2}
\textbf{Thermalization in classical vs quantum mechanics}. \textbf{a}, In classical mechanics, time evolution constructs the thermal state from an initial state that generally bears no resemblance to the former. \textbf{b}, In quantum mechanics, according to the eigenstate thermalization hypothesis, every eigenstate of the Hamiltonian always implicitly contains a thermal state. The coherence between the eigenstates initially hides it, but time dynamics reveals it through dephasing.
}
\end{figure}

J. M. Deutsch and M. Srednicki have independently proposed the scenario (iii), dubbed the
\begin{itemize}
\item[]
\underline{Eigenstate thermalization hypothesis
(ETH)}
[Deutsch\cite{deutsch91} (1991), Srednicki\cite{srednicki94} (1994)].\\
The expectation value $\AEigenv\widehat{A}\Eigenvleft$ of a few-body observable $\widehat{A}$ in an eigenstate of the Hamiltonian $\Eigenvleft$, with energy $E_{\alpha}$, of a large interacting many-body system equals the thermal (microcanonical in our case) average $\langle A \rangle_{\textrm{microcan.}}(E_{\alpha})$ of $\widehat{A}$ at the mean energy $E_{\alpha}$:
\begin{equation}
\AEigenv \widehat{A} \Eigenvleft = \langle A
\rangle_{\textrm{microcan.}}(E_{\alpha}) .
\label{main_eigenstate_thermalization}
\end{equation}
\end{itemize}

The ETH suggests that classical and quantum thermal states have very different natures, as depicted in Fig.\ \ref{fig2}. While at present there are no general theoretical arguments supporting the ETH, some results do exist for restricted classes of systems. To begin with, Deutsch \cite{deutsch91} showed that the ETH holds in the case of an integrable Hamiltonian weakly perturbed by a single matrix taken from a random Gaussian ensemble. Next, nuclear shell model calculations have shown that individual wavefunctions reproduce thermodynamic predictions \cite{horoi95}. Then there are rigorous proofs that some particular quantum systems, whose classical counterparts are chaotic, satisfy the ETH in the semiclassical limit \cite{shnirelman74,voros79,colin85,zelditch87}. More generally, for low density billiards in the semiclassical regime, the ETH follows from Berry's conjecture \cite{srednicki94,heller07}, which in turn is believed to hold in semiclassical classically-chaotic systems \cite{berry77}. Finally, at the other end of the chaos-integrability continuum, in systems solvable by Bethe ansatz, observables are smooth functions of the integrals of motion. This allows the construction of single energy eigenstates that reproduce thermal predictions \cite{korepin93}.

\begin{figure}[!htb]
\begin{center}
\includegraphics[scale=.5]{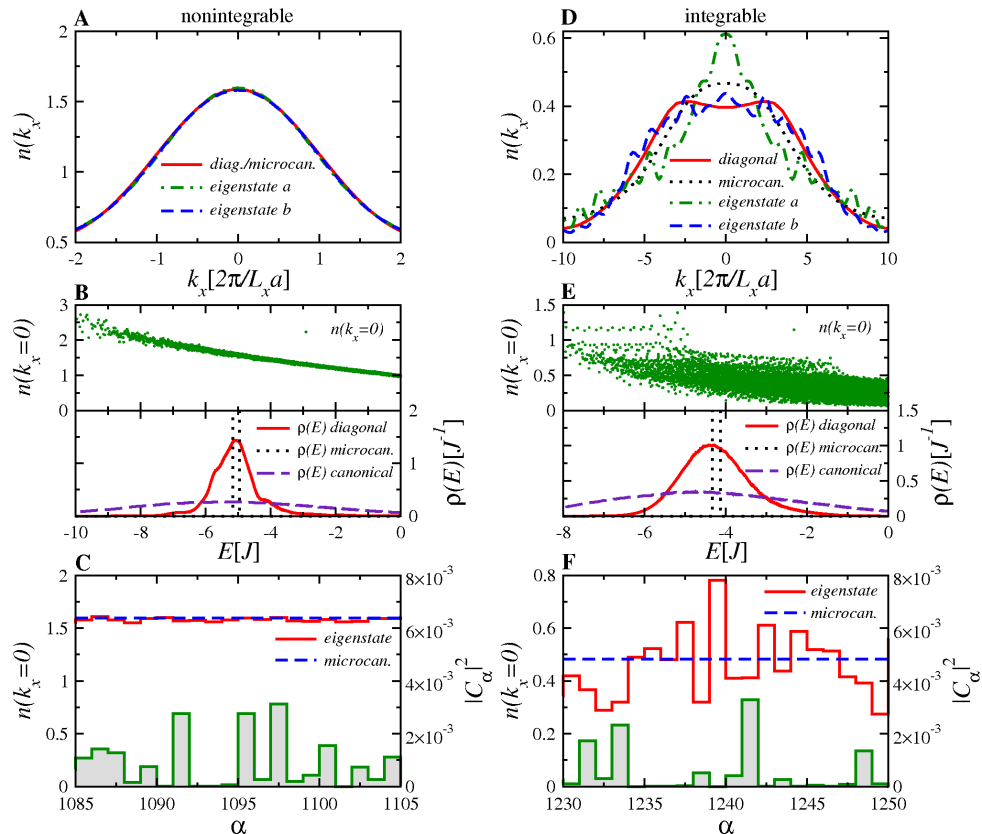}
\end{center}
\vspace{-0.7cm}
\caption {\label{fig3}
\textbf{Eigenstate thermalization hypothesis}. \textbf{a}, In our nonintegrable system, the momentum distribution $n(k_x)$ for two typical eigenstates with energies close to $E_0$ is identical to the microcanonical result, in accordance with the ETH. \textbf{b}, Upper panel: $n(k_x=0)$ eigenstate expectation values as a function of the eigenstate energy resemble a smooth curve. Lower panel: the energy distribution $\rho(E)$ of the three ensembles considered in this work. \textbf{c}, Detailed view of $n(k_x=0)$ (left labels) and $|C_\alpha|^2$ (right labels) for 20 eigenstates around $E_0$. \textbf{d}, In the integrable system, $n(k_x)$ for two eigenstates with energies close to $E_0$ and for the microcanonical and diagonal ensembles are very different from each other, i.e., the ETH fails. \textbf{e}, Upper panel: $n(k_x=0)$ eigenstate expectation value considered as a function of the eigenstate energy gives a thick cloud of points rather than resembling a smooth curve. Lower panel: energy distributions in the integrable system are similar to the nonintegrable ones depicted in \textbf{b}. \textbf{f}, Correlation between $n(k_x=0)$ and $|C_\alpha|^2$ for 20 eigenstates around $E_0$. It explains why in \textbf{d} the microcanonical prediction for $n(k_x=0)$ is larger than the diagonal one.
}
\end{figure}

In Figs.\ \ref{fig3}a-c we demonstrate that the ETH \emphas{is} in fact the mechanism responsible for thermal behavior in our nonintegrable system. Fig.\ \ref{fig3}c  additionally shows that scenario (ii) mentioned above plays no role, because the fluctuations in the eigenstate occupation numbers $|C_{\alpha}|^{2}$ are large. Thermal behavior also requires that both the diagonal and the chosen thermal ensemble have sufficiently narrow energy distributions $\rho(E)$ [$\,=$\ probability distribution $\times$ the density of states], so that in the energy region where the energy distributions $\rho(E)$ are appreciable, the derivative of the curve eigenstate expectation value $A_{\alpha\alpha}$ vs the energy (here $n(k_x=0)$ vs the energy) does not change much; see Supplementary Discussion. As shown in Fig.\ \ref{fig3}b, this holds for the microcanonical and diagonal ensembles but not for the canonical ensemble, explaining the failure of the latter to describe the relaxation in Fig.\ \ref{fig1}. Note that the fluctuations of the eigenstate occupation numbers $|C_{\alpha}|^{2}$ in Fig.\ \ref{fig3}b are lowered by the averaging involved in the computation of the density of states (compare with Fig.\ \ref{fig3}c).

To strengthen the case for the ETH, we tested another observable. We chose it with the following consideration in mind: in our system interactions are local in space, and momentum distribution is a global, approximately spatially additive property. Thus one might wonder if the ETH for momentum distribution arises through some spatial averaging mechanism (we thank the anonymous referee 2 for bringing this point to our attention). It does not: for our final test of the ETH we chose an observable that is manifestly local in space, the expectation value of the occupation of the central site of the lattice. We again found that the ETH holds true (3\% relative standard deviation of eigenstate-to-eigenstate fluctuations).

On the other hand, Figs. \ref{fig3}d-f show how the ETH \emphas{fails} for an isolated one-dimensional \emphas{integrable} system. The latter consists of five hard-core bosons initially prepared in their ground state in an 8-site chain, one of the ends of which we then link to one of the ends of an adjoining (empty) 13-site chain to trigger relaxation dynamics. As Fig.\ \ref{fig3}e shows, $n(k_x)$ as a function of energy is a broad cloud of points, meaning that the ETH is not valid; Fig.\ \ref{fig3}f shows that scenario (ii) does not hold either.

\onecolumngrid

\twocolumngrid

Nevertheless, one may still wonder if in this case scenario (i) might hold---if the averages over the diagonal and the microcanonical energy distributions shown in Fig.\ \ref{fig3}e might agree. Figure \ref{fig3}d shows that this does not happen. This is so because, as shown in Fig.\ \ref{fig3}f, the values of $n(k_x=0)$ for the most-occupied states in the diagonal ensemble (the largest values of eigenstate occupation numbers $|C_\alpha|^2$) are always smaller than the microcanonical prediction, and those of the least-occupied states, always larger. Hence, the usual thermal predictions fail because the correlations between the values of $n(k_x=0)$ and  $|C_\alpha|^2$ preclude unbiased sampling of the latter by the former. These correlations have their origin in the nontrivial integrals of motion that make the system integrable and that enter the \emphas{generalized} Gibbs ensemble, which was introduced in Ref.\ \cite{rigol07} as appropriate for formulating statistical mechanics of isolated integrable systems. In the nonintegrable case shown in Fig.\ \ref{fig3}c, $n(k_x=0)$ is so narrowly distributed that it does not matter whether or not it is correlated with $|C_\alpha|^2$ (we have in fact seen no correlations in the nonintegrable case). 

\begin{figure}[!t]
\begin{center}
\includegraphics[scale=0.94]{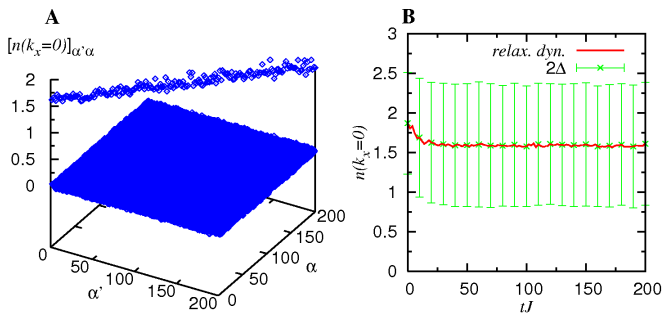}
\end{center}
\caption { \label{fig4}
\textbf{Temporal vs quantum fluctuations}.
\textbf{a}, Matrix elements of the observable of interest, $n(k_{x}=0)$, as a function of state indices; the eigenstates of the Hamiltonian are indexed in the order of diminishing overlap with the initial state. The dominance of the diagonal matrix elements is apparent.
\textbf{b}, The same time evolution as in Fig. \ref{fig1}b  with the error bars showing the quantum fluctuations $n(k_{x}=0)\pm \Delta$ with $\Delta=[\langle \widehat{n}^2(k_x=0)\rangle-\langle \widehat{n}(k_x=0)\rangle^2]^{1/2}$, which are clearly much larger than the temporal fluctuations of $n(k_{x}=0)$.
}
\end{figure}

The thermalization mechanism outlined thus far explains why long-time averages converge to their thermal predictions. A striking aspect of Fig. \ref{fig1}b, however, is that the time fluctuations are so small that after relaxation the thermal prediction works well at every instant of time. Looking at Eq.\ (\ref{timeevolution}), one might think this is so because the contribution of the off-diagonal terms gets attenuated by temporal dephasing, which results from the generic incommensurability of the frequencies of the oscillating exponentials. However, this attenuation only scales as the root of the number of dephasing terms, and is exactly compensated by their larger number: if the number of eigenstates that have a significant overlap with the initial state is $N_{\textrm{states}}$, then typical $C_{\alpha}\sim 1/\sqrt{N_{\textrm{states}}}$, and the sum over off-diagonal terms in Eq.\ (\ref{timeevolution}) finally does not scale down with $N_{\textrm{states}}$:
\begin{equation}
\sum_{\substack{\leftindex,\,\rightindex \\ \leftindex\neq
\rightindex}}\frac{e^{i(E_{\leftindex}-E_{\rightindex})t}}{N_{\textrm{states}}}
A_{\leftindex\rightindex} \sim
\frac{\sqrt{N_{\textrm{states}}^{2}}}{N_{\textrm{states}}}A_{\leftindex\rightindex}^{\text{typical}}
\sim A_{\leftindex\rightindex}^{\text{typical}}
\end{equation}
Hence, were the magnitude of the diagonal and off-diagonal terms comparable, their contributions would be comparable as well, and time fluctuations of the average would be of the order of the average. However, this is not the case and thus
\begin{eqnarray}
A_{\stackrel{\scriptstyle \leftindex\rightindex}{\leftindex\neq \rightindex}}^{\text{typical}} \ll
A_{\alpha^{}\alpha^{}}^{\text{typical}}.
\label{small_off-diagonal}
\end{eqnarray}
Figure \ref{fig4}a confirms this inequality for the matrix elements of the momentum distribution in our system. We should mention that there is an \emphas{a priori} argument---admittedly in part dependent on certain hypotheses about chaos in quantum billiards---in support of this inequality for the case when the mean value of $\widehat{A}$ in an energy eigenstate is comparable to the quantum fluctuation of $\widehat{A}$ in that state \cite{srednicki94a}.

On the other hand, the thermalization we see appears to be working a bit \emphas{too well}: in a system as small as ours, one would expect measurement-to-measurement fluctuations to be much larger than what Fig.\ \ref{fig1}b suggests. Indeed, as we show in Figure \ref{fig4}b, the fluctuations that one would actually measure would be dominated by the quantum fluctuations of the time-dependent state. The rather large size of the quantum fluctuations relative to the thermal mean value is of course particular to small systems; however, the dominance of the quantum fluctuations over the temporal fluctuations of quantum expectation values is not and is actually expected for generic systems in the thermodynamic limit \cite{srednicki96}.

We have demonstrated that, in contrast to the integrable case, the nonequilibrium dynamics of a generic isolated quantum system does lead to standard thermalization. We verified that this happens through the eigenstate thermalization mechanism, a scenario J. M. Deutsch \cite{deutsch91} demonstrated for the case of  an integrable quantum Hamiltonian weakly perturbed by a single matrix taken from a random Gaussian ensemble and M. Srednicki \cite{srednicki94} compellingly defended for the case of rarefied semiclassical quantum billiards, and which both authors conjectured to be valid in general. Our results, when combined with the others we mentioned \cite{deutsch91,horoi95,shnirelman74,voros79,
colin85,zelditch87,srednicki94,heller07,berry77,korepin93}, constitute strong evidence that eigenstate thermalization indeed generally underlies thermal relaxation in isolated quantum systems. Therefore, to understand the existence of universal thermal time-asymptotic states, one should study operator expectation values in individual eigenstates. This is a problem that is linear, time-independent, and conceptually far simpler than any arising in the research---currently dominating the field---on the nonlinear dynamics of semiclassical systems. Among the fundamental open problems of statistical mechanics that could benefit from the linear time-independent perspective are the nature of irreversibility, the existence of a KAM-like threshold \cite{jose98} in quantum systems, and the role of conserved quantities in the approach to equilibrium. Finally, having a clear conceptual picture for the origins of thermalization may make it possible to engineer new, ``unthermalizable'' states of matter \cite{deutsch91}, with further applications in quantum information and precision measurement.

\begin{acknowledgments}
We thank A. C. Cassidy, K. Jacobs, A. P. Young, and E. J. Heller for helpful comments. We acknowledge financial support from the National Science Foundation and the Office of Naval Research. We are grateful to the USC HPCC center where our numerical computations were performed.
\end{acknowledgments}




\section{SUPPLEMENTARY DISCUSSION}

\subsection{1. The Hamiltonian and the numerical calculations.}
In a system of units where $\hbar=1$ the Hamiltonian reads
\begin{equation}
\widehat{H} = -J \sum_{\langle i,j \rangle}
\left(\hat{b}^{\dagger}_i \hat{b}^{}_j + \mbox{h.c.} \right)
+ U \sum_{\langle i,j \rangle} \hat{n}_i\hat{n}_j
\label{HCB_Hamiltonian}
\end{equation}
where $\langle i,j \rangle$ indicates that the sums run over
all nearest-neighbor pairs of sites, $J$ is the hopping
parameter, and $U$ the nearest-neighbor repulsion parameter
that we always set to $0.1J$. The hard-core boson creation
($\hat{b}^{\dagger}_i$) and annihilation ($\hat{b}^{}_j$)
operators commute on different sites,
$[\hat{b}^{}_i,\hat{b}^{\dagger}_j] =\
[\hat{b}^{}_i,\hat{b}^{}_j] =\
[\hat{b}^{\dagger}_i,\hat{b}^{\dagger}_j] =\ 0$ for all $i$ and
$j \ne i$, while the hard-core condition imposes the canonical
anticommutation relations on the same site, $\{ \hat{b}^{}_i,\,
\hat{b}^{\dagger}_i \} = 1$, and
$(\hat{b}^{}_i)^2=(\hat{b}^{\dagger}_i)^2 =  0 $ for all $i$.
Here $\hat{n}_i = \hat{b}^{\dagger}_i \hat{b}^{}_i$ is the
density operator.

\begin{figure}[htb!]
\begin{center}
\includegraphics[scale=.39]{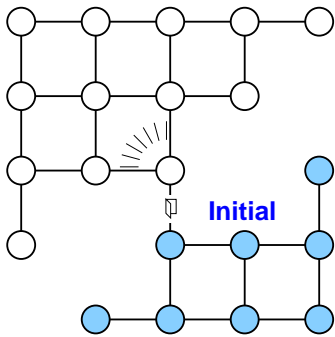}
\end{center}
\vspace{-0.3cm}
\caption{
\label{figSD1}
\textbf{The lattice for the dynamics}. Two-dimensional lattice on which the particles propagate in time. The initial state is the ground state of 5 hard-core bosons confined to the sub-lattice in the lower right-hand corner, and the time evolution starts after the opening of the link indicated by the door symbol.
}
\end{figure}

An exact study of the nonequilibrium dynamics for \emphas{all}
time scales requires a full diagonalization of the many-body
Hamiltonian (\ref{HCB_Hamiltonian}). We are able to fully
diagonalize---essentially to machine precision---matrices of
dimension $D\sim 20,000$, and so we consider $N=5$ hard-core
bosons on a $5\times 5$ lattice with four sites missing
($D=20,349$); see Fig.~\ref{figSD1}. All the eigenstates of the
Hamiltonian are used for the time evolution
\[
 \WFtime=\exp{[-i\widehat{H}t]}\WFini= \sum_\alpha C_{\alpha}\exp{[-i E_\alpha t]}\Eigenvleft \, ,
\]
where $\WFtime$ is the time-evolving state, $\WFini$ the
initial state, $\Eigenvleft$ the eigenstates of the Hamiltonian with the energies $E_\alpha$, and $C_{\alpha}=\langle
\Psi_\alpha | \psi(0) \rangle$. Our initial state is the ground state of the five bosons when they are confined to the lower
part of the lattice (the colored part in Fig.~\ref{figSD1}. The
time evolution begins with the opening of the link shown in
Fig.~\ref{figSD1}, which allows the bosons to expand over the whole
lattice. The position of the missing sites was chosen so that
we only open a single link to start the relaxation dynamics.
The motivation for this will become apparent in the last
paragraph below.

As the principal observables of interest we chose the marginal
momentum distribution along the horizontal axis $n(k_{x}) =
\sum_{k_{y}} n(k_{x},k_{y})$ and in particular its central
component $n(k_{x}=$~$0)$, quantities readily measurable in
actual experiments with ultracold quantum gases \cite{mandel03}. Here
the full two-dimensional momentum distribution is
$n(k_{x},k_{y})=1/L^2\sum_{i,j} e^{-i 2\pi \textbf{k}(\textbf{r}_i-\textbf{r}_j)/L} \langle
\hat{b}^{\dagger}_i\hat{b}^{}_j\rangle$, where
$L=L_{x}=L_{y}=5$ are the linear sizes of the lattice. The
position $\textbf{r}_i=( i_{x} \, d\, , i_{y} \, d)$ involves the
lattice constant $d$.

\subsection{2. The microcanonical ensemble in a small system.}
\begin{figure}[htb!]
\begin{center}
\includegraphics[scale=.7]{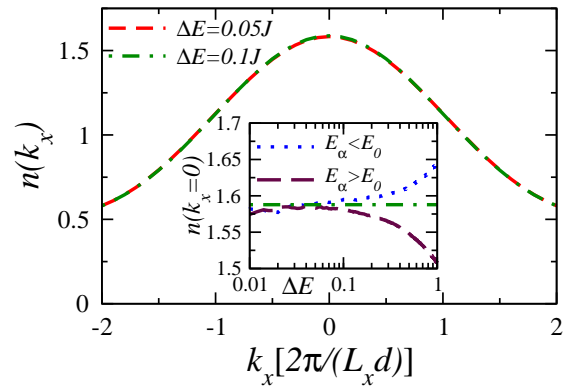}
\end{center}
\vspace{-0.7cm}
\caption{ \label{figSD2}
\textbf{Microcanonical ensemble}. Microcanonical momentum
distribution function for two different values of $\Delta E$.
Inset: Microcanonical predictions for $n(k_x=0)$ calculated
using the left ($[E_0-\Delta E, E_0]$) and the right ($[E_0,
E_0+\Delta E]$) averages as functions of $\Delta E$.
}
\end{figure}

To compute the microcanonical ensemble predictions, we have
averaged over all eigenstates whose energies lie within a
narrow window $[E_0-\Delta E, E_0+\Delta E]$, with
$E_0\equiv\langle \psi(0) | \widehat{H} | \psi(0)
\rangle=\langle \psi(t) | \widehat{H} | \psi(t) \rangle=-5.06 J$. Since
our systems are small there is generally no meaning to the
limit $\Delta E \to 0$, because small enough windows may fail
to contain even a single eigenstate. Instead, one should show
that the microcanonical predictions are robust with respect to
the choice of the width of the energy window. In Fig.~\ref{figSD2}
we demonstrate this robustness in a neighborhood of $\Delta
E=0.1J$, a value that seems to be an appropriate choice given
the data presented in the inset of the same figure. There we
show the dependence on $\Delta E$ of the predictions for
$n(k_x=0)$ given by the ``left-averaged'' and the
``right-averaged'' microcanonical ensembles, by which we mean
that the microcanonical windows are chosen as $[E_0-\Delta E,
E_0]$ and $[E_0, E_0+\Delta E]$, respectively. We see that for
$\Delta E\lesssim 0.1J$, the two microcanonical predictions are
almost independent of the value of $\Delta E$. The main panel
in Fig.~\ref{figSD2} shows that the microcanonical values of
$n(k_x)$ for $\Delta E=0.05J$ and for $\Delta E=0.1J$ are
indistinguishable.\\

\subsection{3. Eigenstate thermalization and the width of the
energy distribution.} The eigenstate thermalization alone is
not sufficient to ensure an agreement between the predictions
of the diagonal and thermal ensembles. As discussed in Ref. 
\cite{srednicki94}, it is also necessary that both distributions be sufficiently
narrow. More specifically, one must require for both ensembles
\begin{equation}
(\Delta E)^2\, |A''(E)/A(E)| \ll 1,
\label{narrowness_criterion}
\end{equation}
where $\Delta E$ is the width of the energy distribution in the
ensemble, and $A(E)$ is the dependence of the expectation value
of the observable $A_{\alpha\alpha}=\AEigenv \widehat{A}
\Eigenvleft$ on the energy $E_{\alpha}$ of the Hamiltonian-operator
eigenstate $\Eigenvleft$. Note that because of eigenstate
thermalization, $A(E)$ is a smooth function of energy. For the
thermodynamical ensembles the condition
(\ref{narrowness_criterion}) is always satisfied in the
thermodynamic limit. We now show that it is also satisfied for
the diagonal ensemble in the thermodynamic limit.

If one considers an observable $a$ that is the intensive
counterpart of $A$, all conclusions obtained for $a$ can be
extended to the original observable $A$ via trivial rescaling.
For example, for our principal observable of interest,
$n(k_{x})$, the corresponding intensive variable is the
momentum density $\xi(p_{x})$ normalized as $\int \!dp_{x}\, \xi(p_{x}) = 1$. Notice that in this case $\xi(p_{x}) =
n(k_{x}) L_{x} d/(2\pi N)$.

For $a$, the condition in (\ref{narrowness_criterion}) reduces
to
\begin{equation}
(\Delta \epsilon)^2 |a''(\epsilon)/a(\epsilon)| \ll 1,
\label{narrowness_criterion_2}
\end{equation}
where $\epsilon \equiv E/N$. For sufficiently large systems the
dependence of $a$ on $\epsilon$ is independent of the system
size. Hence, in order to justify the validity of
(\ref{narrowness_criterion_2}) it is sufficient to prove that
the width of the distribution of the energy per particle in the
diagonal ensemble converges to zero for large linear sizes $L$
of the system:
\begin{equation}
\Delta \epsilon \stackrel{L\to\infty}{\longrightarrow} 0 \, .
\label{narrowness_criterion_3}
\end{equation}

Suppose that initially our system is prepared in an eigenstate
$|\Psi_0\rangle$ of a Hamiltonian $\widehat{H}_{0}$ and that at
time $t=0$ the Hamiltonian is suddenly changed to
$\widehat{H}$:
\[
\widehat{H}_0 \to \widehat{H} = \widehat{H}_{0} + \widehat{W},
\]
where $\widehat{W}$ is the difference between the new and the
old Hamiltonians. Within this scenario, the energy width
\[
 \Delta E = \sqrt{\sum_\alpha E^2_\alpha |C_{\alpha}|^{2}
          - \left( \sum_\alpha E_\alpha |C_{\alpha}|^{2} \right)^2 }
\]
of the diagonal ensemble becomes equal to the variance of the
new energy in the state $|\Psi_0\rangle$:
\[
\Delta E = \Delta H \equiv \sqrt{\langle \Psi_0 | \widehat{H}^2 | \Psi_0 \rangle
-\langle \Psi_0 | \widehat{H} | \Psi_0 \rangle^2
}\, .
\]
It is now straightforward to show that the variance of
$\widehat{H}$ equals the variance of $\widehat{W}$:
\[
\Delta H = \Delta W.
\]

In order to deduce how $\Delta W$ scales in the thermodynamic
limit, we assume that $\widehat{W}$ is a sum of local operators
$\hat{w}(j)$ over some region of the lattice $\sigma$ (a single
point, a straight line, the whole lattice, etc.):
\[
\widehat{W} = \sum_{j \in \sigma} \hat{w}(j).
\]
Here $\hat{w}(j)$ is a polynomial of creation and annihilation
operators localized at the points $j+\Delta j$, where $|\Delta
j|$ is limited from the above by a finite number that does not
scale with the system size. The mean square of $\widehat{W}$
can be written as
\begin{multline}
\label{Delta_W}
\langle \Psi_0 | \widehat{W}^2 | \Psi_0 \rangle =
\langle \Psi_0 | \widehat{W} | \Psi_0 \rangle^2 \\
+ \sum_{j_1,j_2 \in \sigma}
\left[
  \langle \Psi_0 | \hat{w}(j_1)\hat{w}(j_2)
| \Psi_0 \rangle \right. \\
- \left.
  \langle \Psi_0 | \hat{w}(j_1)
| \Psi_0 \rangle
  \langle \Psi_0 | \hat{w}(j_2)
| \Psi_0 \rangle
\right ].
\end{multline}

In the absence of long-range correlations the expression in
brackets tends to zero for large distances between $j_1$ and
$j_2$. Therefore, the whole second term on the right-hand-side
of (\ref{Delta_W}) scales as $L^{d_{\sigma}}$, where
$d_{\sigma}$ is the dimensionality of the sublattice $\sigma$
and $L$ is the linear size of the lattice. The variance of
$\widehat{W}$ scales the same way:
\[
(\Delta W)^2 \stackrel{L\to\infty}{\propto} L^{d_{\sigma}}.
\]
Retracing our steps, we arrive at the conclusion that the
energy width $\Delta \epsilon$ indeed tends to zero in the
thermodynamic limit:
\[
\Delta \epsilon  \stackrel{L\to\infty}{\propto} \frac{1}{L^{d_L-d_{\sigma}/2}},
\]
where $d_L \ge d_{\sigma}$ is the dimensionality of the whole
lattice.

Note that for the two-dimensional lattice considered in this
paper the role of $\widehat{W}$ is played by the hopping energy
of the ``door'' link. An analysis similar to the one above
shows that increasing the number of ``door'' links will lead to
an increase in the width $\Delta \epsilon$, proportional to the
square root of the number of ``door'' links. This is why in our
2D experiment, we have chosen the position of the missing sites
to be the one in Fig.~\ref{figSD1}, so that only a single link is
opened during the time evolution.


\end{document}